# Mineralogical and Thermal Analyses of a Bangle Shard from Harrappa, an Indus Valley Settlement in Pakistan

Saheeb Ahmed Kayani[1], Rehan ul Haq Siddiqui[2]

[1]National University of Sciences and Technology, Islamabad, Pakistan,
[2]Geoscience Advance Research Laboratories (Geological Survey of Pakistan), Islamabad, Pakistan
[1]saheebk@ceme.nust.edu.pk

**Abstract**

In this research study we present initial results of a recent project in which mineralogical and thermal analysis were carried out on a terracotta bangle shard from Harrappa. We were surprised to find bentonite clay as the major constituent of the bangle shard. Also we have used knowledge of the mineralogical structure determined through X-ray diffraction and results of thermal analysis to predict value of firing temperature of the bangle shard.

## 1. Introduction

In Pakistan, Harrappa is recognized as one of the most important and magnificent site of the Indus Valley Culture that flourished around 2600 BCE in this region. Research studies carried out at this settlement have provided wealth of information about its people and their life style. One fascinating aspect of Harrappa is the abundance and variety of pottery that has been unearthed at this site. This includes every day utensils, storage tanks for grain, toys for children, trading seals, figurines, and terracotta jewellery including bangles, beads, and necklaces.

The ancient potters and artisans have displayed great ingenuity in creating forms and styles of jewellery that were appealing to the eye and at the same time light in weight and tough, making them easy to wear and lasting. Given the nature of use of these ornaments, it is likely that the potters employed different or especially processed raw materials in producing them. In order to determine the type of contents used in making bangles in Harrappa, a specimen shard (Fig. 1) was tested using the facilities available in Geoscience Advance Research Laboratories (a facility of Geological Survey of Pakistan) in Islamabad.

Fig. 1: Bangle shard photographed with Pakistani one rupee coin as reference. Diameter: 0.95-1 cm, length: 5.65 cm or about 2/5th of the complete bangle.

The mineralogical make up of the clay used in the bangle shard was determined using XRD analysis while simultaneous TG/DT analyses were carried out to establish thermal properties and nature of phase changes that took place while the bangle was processed. (XRD analysis was carried out on 'Philips X'Pert PRO X-ray diffractometer' and thermal analysis (TG/DTA) was carried out on 'Shimadzu DTG-60H differential thermal and thermogravimetric analyser'.)

## 2. Results and Discussion

XRD analysis shows that the mineralogical make up of the clay used in the bangle shard includes montmorillonite $((Na,Ca)_{0.33}(Al,Mg)_2(Si_4O_{10})(OH)_2 \cdot nH_2O)$, gypsum $(CaSO_4 \cdot 2H_2O)$, and quartz $(SiO_2)$. Results of thermal analysis have been included in Fig. 2. The temperature was varied in steps of 20°C/min from room temperature to 1200°C. For a sample mass of 20.32 mg, a loss of 1.471 mg or 7.239% was observed. There is a regular trend in mass loss over the entire heating range, which shows that the clay or clays used in the bangle shard are almost uniform in composition and properties. Mass loss in clays during thermogravimetric analysis can be divided into three stages: dehydration from room temperature to about 200°C, decomposition of hydroxyls from 400°C to 650°C, and decomposition of carbonates in the range 700-800°C. [1]

Fig. 2: Results of simultaneous TG/DT analyses.

The TGA curve for the specimen shows mass loss occurring in three distinct steps. The initial mass loss (0.15 mg) attributed to dehydration is very small, but it coincides with an endothermic peak at 139.16°C on DTA curve. (An endothermic effect attributed to presence of gypsum is also observed in the range 100-120°C. [2]) From this temperature onwards there is a gradual mass loss of 0.75 mg up till 850°C. Finally a mass loss of 0.5 mg is recorded from 850°C to 1200°C on the last portion of TGA curve.

The mass loss around 400-650°C which coincides with a sharp exothermic peak at 546.5°C on DTA curve can be attributed to combustion of organic material in the tested specimen. [2] This organic material may have been added by ancient potters as a binder. (From XRD analysis, abundance of montmorillonite and the characteristic two endothermic peaks on DTA curve provide the evidence that the major constituent of the tested specimen is bentonite. The thermal analysis curves for the tested specimen can be compared with modern industrial bentonite (from India) and the trends are almost similar. [3] A difference (in peaks and ranges) may exist due to addition of other components in the raw ceramic material of the bangle shard specimen.) Studies have shown that montmorillonite is destroyed when fired to a temperature of 860°C. [1] If montmorillonite has survived the firing stage during processing of the bangle, it appears that the temperature in the firing kiln was not in excess of 860°C. As progress in pottery making is related with firing temperature (the higher the firing temperature, the more advanced the technique) it is evident that the skills of ancient Harrappan potters were well developed and fairly advanced for their time. [4]

Quartz has also been detected in XRD analysis. It is a non-plastic material where as bentonite is known to be very plastic and suited for complex shapes. Quartz may have been added in the raw ceramic material as a temper to allow water to evaporate smoothly (thus avoiding cracking) and also to improve handling and working of the raw clay. But unknown to ancient potters of Harrappa, this addition may have caused the bangles made out of this raw material to become less resistant to mechanical stresses while in use.

An endothermic effect is observed from 600-950°C. There is a sharp endothermic peak at 993.69°C followed by an exothermic effect at about 1020°C. The endothermic effect especially around 800-950°C can be attributed to decomposition of any carbonates in the raw ceramic material used for making this particular bangle. No amount of calcite or dolomite has been identified in XRD analysis (given the limitation of the XRD apparatus not being able to detect minerals less than 10 wt% of the sample), the most usual carbonates in ancient ceramic materials. [2] The endothermic peak at 993.69°C can be interpreted as an indication of appearance of a crystalline phase (i.e. due to vitrification of quartz) in the sample (above 800°C), which might have undergone a solid-phase polymorphic transformation at this high temperature. [5][2]

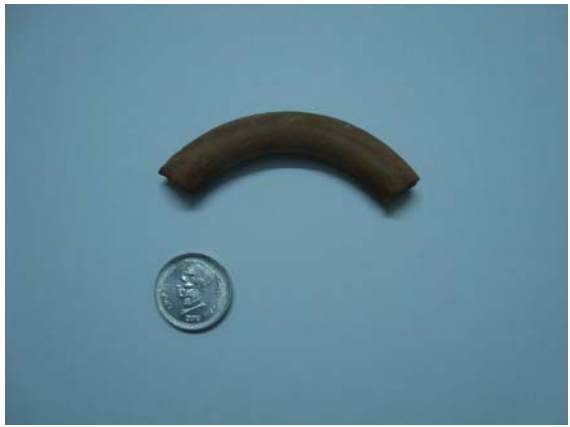

**Figure 1**

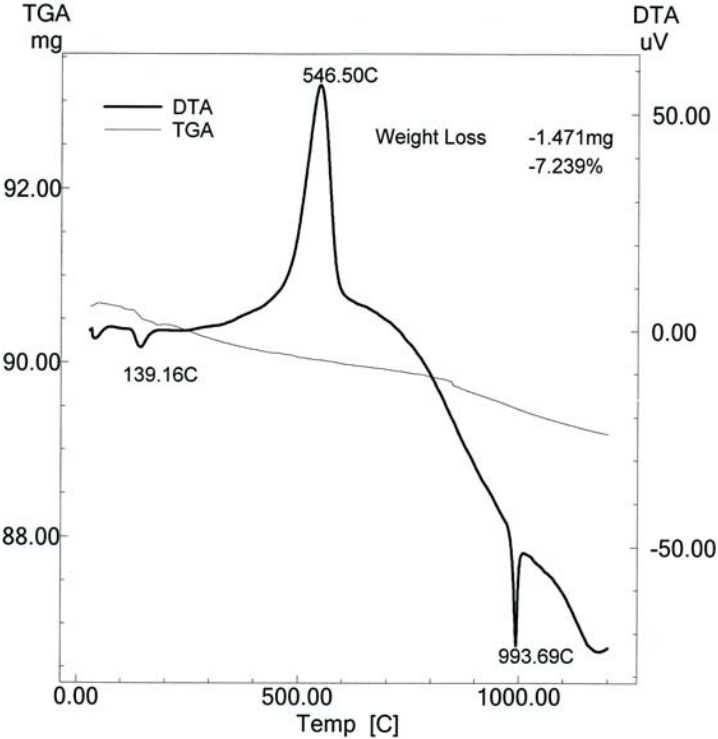

**Figure 2**